\documentstyle{cupconf}
\input epsf

\ifoldfss
\else
  \ifnfssone
    \newmathalphabet{\mathit}
      \addtoversion{normal}{\mathit}{cmr}{m}{it}
      \addtoversion{bold}{\mathit}{cmr}{bx}{it}
    \newmathalphabet{\mathcal}
      \addtoversion{normal}{\mathcal}{cmsy}{m}{n}
    \else
    \ifnfsstwo
    \fi
  \fi
\fi

\def\Rey{\mbox{\it Re}}   
\def\hexnumber#1{\ifcase#1 0\or1\or2\or3\or4\or5\or6\or7\or8\or9\or
 A\or B\or C\or D\or E\or F\fi }

\makeatletter
\ifx\CUP@mtlplain@loaded\undefined
\else
\fi
\makeatother
 \makeatletter
 \ifx\CUP@mtlplain@loaded\undefined
   \font\tenbmi=cmmib10 at 10pt
   \font\sevenbmi=cmmib10 at 7pt
   \font\fivebmi=cmmib10 at 5pt

   \newfam\bmifam
   \textfont\bmifam=\tenbmi
   \scriptfont\bmifam=\sevenbmi
   \scriptscriptfont\bmifam=\fivebmi
   
 \fi
 \makeatother
\ifnfsstwo

\fi
\ifnfssone

\fi
\ifoldfss

\fi

\mathchardef\varLambda="0103

\makeatletter
\ifx\CUP@mtlplain@loaded\undefined
\else
\fi
\makeatother
\makeatletter
\ifx\CUP@mtlplain@loaded\undefined
  \font\tenbms=cmbsy10
  \font\sevenbms=cmbsy10 at 7pt
  \font\fivebms=cmbsy10 at 5pt
  \newfam\bmsfam
  \textfont\bmsfam=\tenbms
  \scriptfont\bmsfam=\sevenbms
  \scriptscriptfont\bmsfam=\fivebms

  \edef\bsy@{\hexnumber\bmsfam}
  \mathchardef\bnabla="0\bsy@72
\fi
\makeatother
\title[Reynolds number of the Reynolds' layer]{What is the Reynolds number of the Reynolds' Layer?\footnotemark}

\author[R. A. Benjamin]%
{R\ls O\ls B\ls E\ls R\ls T\ns A.\ns B\ls E\ls N\ls J\ls A\ls M\ls I\ls N$^1$\thanks{Previous address: Minnesota Supercomputer Institute, 1200 Washington Ave. South, Minneapolis, MN 55415}}

\affiliation{$^1$Department of Physics, University of Wisconsin-Madison,
1150 University Ave, Madison, WI 53706, USA}

\begin{document}
\ifnfssone
\else
  \ifnfsstwo
  \else
    \ifoldfss
      \let\mathcal\cal
      \let\mathrm\rm
      \let\mathsf\sf
    \fi
  \fi
\fi

\maketitle

\begin{abstract}

Several authors have now suggested that some interstellar clouds above
the plane of the Galaxy are interacting with the Reynolds' layer, the
warm ionized gas extending well above ($H \cong 910~pc$) the Galactic
plane (Reynolds 1993).  Characterizing the interaction between these
clouds and their surroundings should be useful in understanding one
source of interstellar turbulence: vertical shear flows. This paper
discusses how studies of the morphology and drag coefficient of
falling clouds might be used to constrain the Reynolds number for the
flow, and hence the effective viscosity of the warm ionized medium. If
arguments based on morphology are correct, the effective viscosity of
the warm ionized medium is significantly higher than the classical
values. Possible resolutions to this problem are suggested.

\end{abstract}

\firstsection 

\section{Turbulence from Vertical Flows}

The spectrum of density and velocity fluctuation in the ionized
interstellar medium (ISM) measured by scintillation of pulsars
suggests that on small scales much of the structure of the diffuse
ionized ISM may arise as the result of turbulent processes. Turbulence
arises in regions of viscous shear flows. In the Galaxy, such flows
have a large range of outer length scales, and include galactic
rotational shear in both the radial and vertical (c.f. Walterbos 1998,
this volume) directions, spiral density waves, stellar mass outflows
(jets, winds, and explosions), and photoionization-driven flows. The
structures formed contain energy over a range of length scales which
is ultimately dissipated via viscous (hydrodynamical) and resistive
(magneto-hydrodynamical) processes.

An additional source of interstellar turbulence is the shear flow
generated by buoyant motions of rising bubbles (Parker 1992) or
falling clouds. Since the sizes and velocities of at least some clouds
are known, they provide a good environment to study one mechanism for the
production of interstellar turbulence and constrain some relevant physical
parameters.

\footnotetext{$^{\dagger}$ Grammar afficianados will note that while objects, 
such as the warm ionized layer, take the possessive form, dimensionless
numbers do not (Vogel 1983).}

\section{How do clouds fall...}

{\it ...At terminal velocity?} Classical dynamics has conditioned us
to think that objects speed us as they fall and that the acceleration
is independent of the mass of the object. But experience shows that
this is not always the case. A hammer, for example, falls faster than
a feather.  This is because of the presence of an external fluid
medium. If the viscosity of the fluid is non-zero, there will be
momentum transferred from the object to the surrounding
fluid. Drag is a tricky thing to calculate, and counterintuitive things 
can happen. For
instance, if the ambient fluid has a density gradient, as in a
gravitationally stratified atmosphere, sufficiently ``light'' objects
will slow down as they fall rather than speed up, because the drag
force decellerates the object to terminal velocity, and drag increases
as the object falls down the density gradient. This effect is seen in
raindrops (Foote \& du Toit 1969).

In Benjamin \& Danly (1997) (BD97), it was suggested that the same effect,
where the Reynolds' layer plays the role of the
Galactic atmosphere, produces a velocity stratification of neutral
hydrogen clouds with height. Measurements of the column density and
velocity of these clouds, along with models of the gravitation field
and external density, were used to predict cloud distances,
which could be checked with absorption lines studies. There was good agreement with the
available data, including correctly predicting the distance to the
high velocity cloud Complex M. {\it It also provides a satisfactory 
explanation for
why high velocity clouds are high velocity, i.e.  they
exist in distant, low drag environment.} Just \underline{how} distant
remains a point of some debate (see Wakker \& van Woerden 1997).

\begin{figure}[tb]
 \centerline{
 \epsfysize= 5.0in
 \epsffile{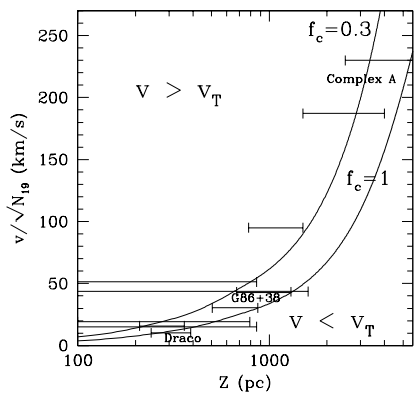}}
\caption{Plot of normalized vertical velocity  of cloud as a function
of cloud distance. The two curves $A(z)$ and $1.8*A(z)$ are the 
normalized terminal velocity predictions for cases with $f_{c}=1$ and $f_{c}=0.3$. The terminal velocity hypothesis agrees well with the data, but
more cloud distances and tighter distance brackets are needed.}
\end{figure}

The terminal velocity is given by $v_{T}(N,z)=\sqrt{2 N_{HI}
g(z)/n_{h}(z)f_{c}C_{D}}$, where $f_{c}$ is the cloud neutral
fraction, and $C_{D}$ is the drag coefficient.
Clouds falling at a terminal velocity obey the relation

\begin{equation}
\frac{v}{\sqrt{N_{HI}}}=\sqrt{\frac{2 g(z)}{n_{h}(z)f_{c}C_{D}}}=A(z)
\end{equation}

which isolates the observable quantities on the left hand side, and
the model assumptions for gravity and gas density on the right. We can
then ask whether the normalized velocity, $v/\sqrt{N_{H~I}}$, of
observed clouds depend upon height. This is shown in Figure 1, which
plots the normalized velocity vs. the distance brackets on
clouds. Most clouds have $|b|> 50^{\circ}$ to minimize the
contribution of galactic rotation to the observed velocity.  The curve
shows a ``best guess'' $A(z)$ from BD97, where the density includes
the H I, warm ionized medium, and a hot halo. We have also added three
cloud with distances determined subsequent to this paper, although
they are at less than $b=50^{\circ}$: Draco at $b=38$, G86.0+38.3 (Gladders
1998), and high velocity cloud complex A at $b=43$ (Wakker 1998).  {\it Note 
that the terminal velocity formula successfully predicted the distance 
to these clouds.} These data show a trend for $v/\sqrt{N}$ to
increase with height. In an ideal situation,
the curve $A(z)$ could be shifted to solve for the best value of
$C_{D}$. Right now, all one can say is that $C_{D}<20$ and is
consistent with being one, expected theoretically (Jones et al 1996).

Without more data, it is hard to be sure whether the
terminal velocity hypothesis is true for some or all of the diffuse
galactic clouds. Our ignorance of how clouds relate to the intercloud
medium is marked.  Other suggestions that can be found in the
literature are the following:

{\it ...Ballistically?} In the originally envisioned galactic fountain
model (Shapiro \& Field 1976; Bregman 1980), clouds condensed out of
thermal instabilities and were assumed to fall ballistically back to
the disk. This assumption was in keeping with the ballistic cloud
models of Oort (1954) and Spitzer (1978). Although I have found that
most people personally disavow the ballistic option, discussion of
cloud ``trajectories'' and ``orbits'' which neglect the effects of
drag continue to be promulgated through the literature.

{\it ...Apart?} The timescale for the Rayleigh-Taylor and
Kelvin-Helmholtz instability to develop in falling clouds is shorter
than the free-fall time. However, such instabilities may be inhibited
by incorporation of radiative cooling (Vietri et al 1996) or magnetic
fields (MacLow et al 1994). This point of view is supported by the
fact that the internal velocity dispersion of high velocity clouds is
sufficiently small that the clouds should not disperse before hitting
the plane.  Note also that raindrops are also subject to the same
instabilities.  In that case, surface tension saves the day (and the
crops).

{\it ... They don't?} There are several possible forces that might act
to prevent a cloud from falling: photolevitation (Franco et al 1991),
galactic winds, or magnetic tension (Franco 1998, this volume). Of
course, if these are always operative, there would be no vertical mass
circulation. So although these mechanisms might act freqently, there
must be times when they break down.

{\it What do you mean ``cloud''?} The ``interstellar cloud'' as a
discrete and time-evolving entity may be a misleading concept (c.f.,
Scalo 1990). If clouds are just short-lived density concentrations in
the interstellar fluid, talking about the time history of an
coherent fluid element may be nonsensical.

\section{Viscosity}

Here, I will take the point of view that clouds \underline{do} fall
due to lack of buoyancy and interact with the intercloud medium. This
point of view can be be supported by the velocity-distance correlation
discussed above, the work of Odenwald (1988) and Reach, Wall, \&
Odegard (1998) who show several Galactic ``cometary'' clouds from IRAS
and DIRBE surveys; Howk \& Savage (1997), who show cometary clouds 1
kpc above the plane of NGC 891; Kerp et al (1996), who argue for a
positional correlation between H I HVC and X-ray emission; Pietz et al
(1996) who show velocity bridges between low and high velocity gas,
and so on. These papers suggest that at least some clouds are being
shaped by interaction with the ambient medium. What physical
information can we glean from these environments?

The fundamental parameter that characterizes a flow of velocity $V$
around an object of length $L$ is the Reynolds number,
$\Rey=LV/\nu_{eff}$, where $\nu_{eff}$ is the ``effective'' kinematic
viscosity. I say ``effective'', because what is called turbulent
viscosity and magnetic effects may and probably do come into play. In
the absence of magnetic fields, the classical kinematic viscosity
${\rm (cm^{2}~s^{-1})}$ of the ISM is $\nu=6 \times 10^{19}
T_{4}^{5/2} n_{-2}^{-1}$. In the presence of a magnetic field,
momentum transfer perpendicular to the direction of the magnetic field
is decreased, and the viscosity decreases by twelve (!) orders of
magnitude to $\nu=2.8 \times 10^{7} B_{3 \mu G}^{-2}
T_{4}^{-1}n_{-2}$. Flows of similar Reynolds numbers (all other
things, like Mach number, being equal) will have similar (1)
morphological characteristics, (2) drag coefficients, and (3) energy
dissipation rates. All three of these quantities may be constrained
observationally, giving an estimate for the Reynolds number. Here, I
concentrate on the Draco molecular cloud. This cloud shows a
``cometary'' morphology (see Odenwald \& Rickard 1987).
Assuming purely vertical motion, its downward velocity is
$V_{z}=-34~km~s^{-1}$ with a column density of $N(H I) \cong 1.8
\times 10^{20}~cm^{-2}$. Using the recently determined distance of
$240 < z < 390 pc$ (Gladders 1998), its size is
$L=3.4-5.4~pc$.  Assuming that $T=8000~K$ and $n_{e}=0.02~cm^{-3}$ 
in the above formulae,
the Reynolds number for the cloud \underline{should} be
$\Rey=2 \times 10^{6}$ or $\Rey=6 \times 10^{17}$, for the viscosity
without or with magnetic fields, respectively. How do our empirical
measurement compare?

{\bf Morphology:} Odenwald (1988) identified 15 IRAS $100~\mu m$
clouds with a ``head-tail'' or cometary appearance, suggesting that
the morphology of these clouds arose from cloud-ISM
interaction. (Interestingly, several of these clouds also showed
evidence of star-formation, possibly triggered by the same
interaction.) Under the assumption that the
morphology of the tail was due to the "wake" produced by the cloud,
Odenwald associated a Reynolds number with the morphological
characteristics of the cloud.  Flows of extremely low \Rey~ (high
viscosity) will have smooth, laminar flow patterns; flows of
intermediate \Rey~ will produce irregular clumps and vortices that
will detach from the back of the cloud; very high \Rey~ will produce
fully turbulent flow, in which the wake will appear smooth
again. Based on an elongated and clumped morphology, Odenwald assigned
the flow around the Draco cloud with $\Rey~ < 50$.

{\bf Drag coefficient:} The recently determined distance to the Draco
cloud matches the terminal velocity prediction of BD97 assuming that
$f_{c}=1$ and provided that $0.9<C_{D}<1.2$.  Unfortunately, no study
exists characterizing how the drag coefficient of interstellar clouds
should vary with Reynolds number. This is partially because clouds,
unlike solid objects, may deform and even fragment depending upon the
initial conditions and constituent physics, both of which are poorly
known.  Numerical simulations of Jones et al 1996, among others,
suggest that $C_{D}=1$ for simulations which have an effective $\Rey~
\approx $ few hundred. Based on flows around solid objects, $C_{D}
\rightarrow 1$ for large enough $\Rey$. Comparing our derived range of
drag coefficient to the value for flow around a sphere (Landau \&
Lifschitz 1987, Fig 34) yields $Re >90$.

\section{The Reynolds number}

The alert reader will have noticed that the two estimates of the 
Reynolds number for the same halo cloud are disjoint. However, there is
sufficient uncertainty in both of them that they may both be 
reconciled. Given that both are estimated using information for solid 
objects moving through a fluid medium, there will need to be some modification
in the above numbers. The possibility of a bow shock will also be important.
However, the difference between the Reynolds number based on morphology
and the classical value differ by more than four orders of magnitude.
Either the morphological interpretation is in error or the viscosity
is significantly higher than the classical value. Both of these are possible, 
but it is difficult to distinguish between them.
Those experienced in accretion disk physics are well aware of the 
possibility of anomalous viscosity; the magneto-rotational instability
discussed by Gammie (1998, this volume) is one mechanism. But even shakier
still is our ability to convert morphlogy into a physical scenario.
For instance, rather than the tail being produced in a turbulent flow,
it could be material ablated from the cloud via a Kelvin-Helmholtz 
instability. Another example is G110-13 (Odenwald et al 1992),
a cometary cloud which turned out to be a probable cloud-cloud 
collision. The
lesson here is {\it interpreting morphology in the absence of other information is dangerous.} Numerical simulations of this process will probably
be needed to build some intuition as to what such processes should
actually look like.

\begin{acknowledgments}
I would like to thank Pepe Franco and the other organizers for a most
original and stimulating meeting and Don Cox (and NASA grant
NAG5-3155) for sending me there.  Some of the work done here was done
using the facilities of the Minnesota Supercomputer Institute.
\end{acknowledgments}

\end{document}